\documentclass[preprint,12pt]{elsarticle}

\usepackage{amssymb}
\usepackage{amsmath}

\usepackage{url}
\usepackage{hyperref}
\usepackage{xcolor}

\begin{document}

\begin{frontmatter}



\title{Recent advances in perturbative QCD at high density}

\author[l1]{Pablo Navarrete\corref{pn}}
\cortext[pn]{pablo.navarrete@helsinki.fi}
\author[l1]{Aleksi Vuorinen\corref{av}}
\cortext[av]{aleksi.vuorinen@helsinki.fi}

\affiliation[l1]{organization={Department of Physics and Helsinki Institute of Physics},
            addressline={P.O.~Box 64},
            postcode={FI-00014 University of Helsinki},
            country={Finland}}



\begin{abstract}
In these conference proceedings, we discuss recent progress in high-order perturbative studies of the thermodynamic and transport properties of dense quark matter. Special emphasis will be placed in the introduction of a promising new computational tool, thermal Loop Tree Duality, which enables pushing the existing weak-coupling calculations to higher perturbative orders.
\end{abstract}


\end{frontmatter}



\section{Introduction}
\label{sec:intro}

With the first successful production of quark–gluon plasma in heavy-ion experiments still a fresh memory \cite{STAR:2005gfr}, considerable attention in the study of Quantum Chromodynamics (QCD) matter has recently shifted to the cores of neutron stars (NSs),  where deconfined matter may exist in a stable form \cite{Baym:2017whm}. This is largely due to a series of landmark observations during the past decade that have turned these extreme astrophysical objects to a precision laboratory of nuclear matter, with the femtometer physics of QCD encoded in electromagnetic (EM) and gravitational-wave (GW) signals. The most important forms of such NS data include radio and X-ray measurements of pulsar masses \cite{Antoniadis:2013pzd,NANOGrav:2019jur} and radii \cite{Miller:2021qha,Riley:2021pdl}, respectively, as well as both GW \cite{LIGOScientific:2017vwq} and EM \cite{LIGOScientific:2017ync} recordings of binary NS mergers.

To turn the influx of novel NS observations to fundamental-physics insights, comparable advances are required on the microphysics side. With the Sign Problem of lattice QCD preventing a direct \textit{ab-initio} approach at densities corresponding to NS cores or a transition from hadronic matter (HM) to quark matter (QM) \cite{deForcrand:2009zkb}, attention turns to first-principles approaches, valid in specific limits. Indeed, for sub-saturation-density nuclear matter, approaches based on a Chiral-Perturbation-Theory power counting have reached the Next-to-Next-to-Leading Order (N2LO) for the Equation of State (EoS) (see, e.g., \cite{Tews:2012fj,Drischler:2017wtt}), while in QM at tens of saturation densities, only one contribution remains missing from the full N3LO perturbative pressure \cite{Gorda:2021znl,Gorda:2023mkk}. As demonstrated in \cite{Komoltsev:2021jzg}, advances in either of these limits would result in quantitative improvements in the model-independent inference of the NS-matter EoS, a very active topic of investigation in modern nuclear astrophysics \cite{Annala:2017llu,Capano:2019eae,Annala:2021gom,Raaijmakers:2021uju,Huth:2021bsp}.

\begin{figure}
    \centering \includegraphics[width=0.51\textwidth]{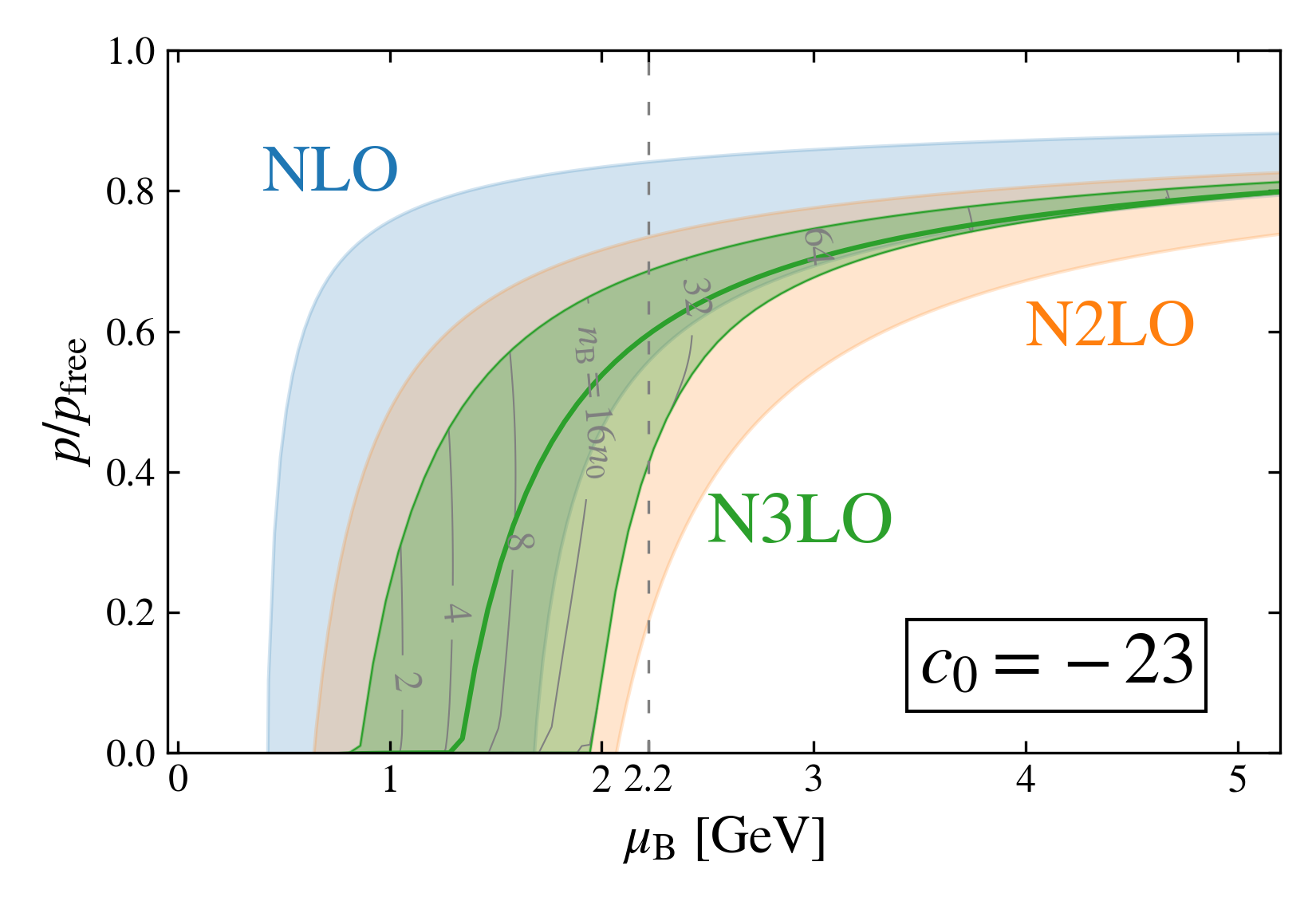}\;\includegraphics[width=0.475\textwidth]{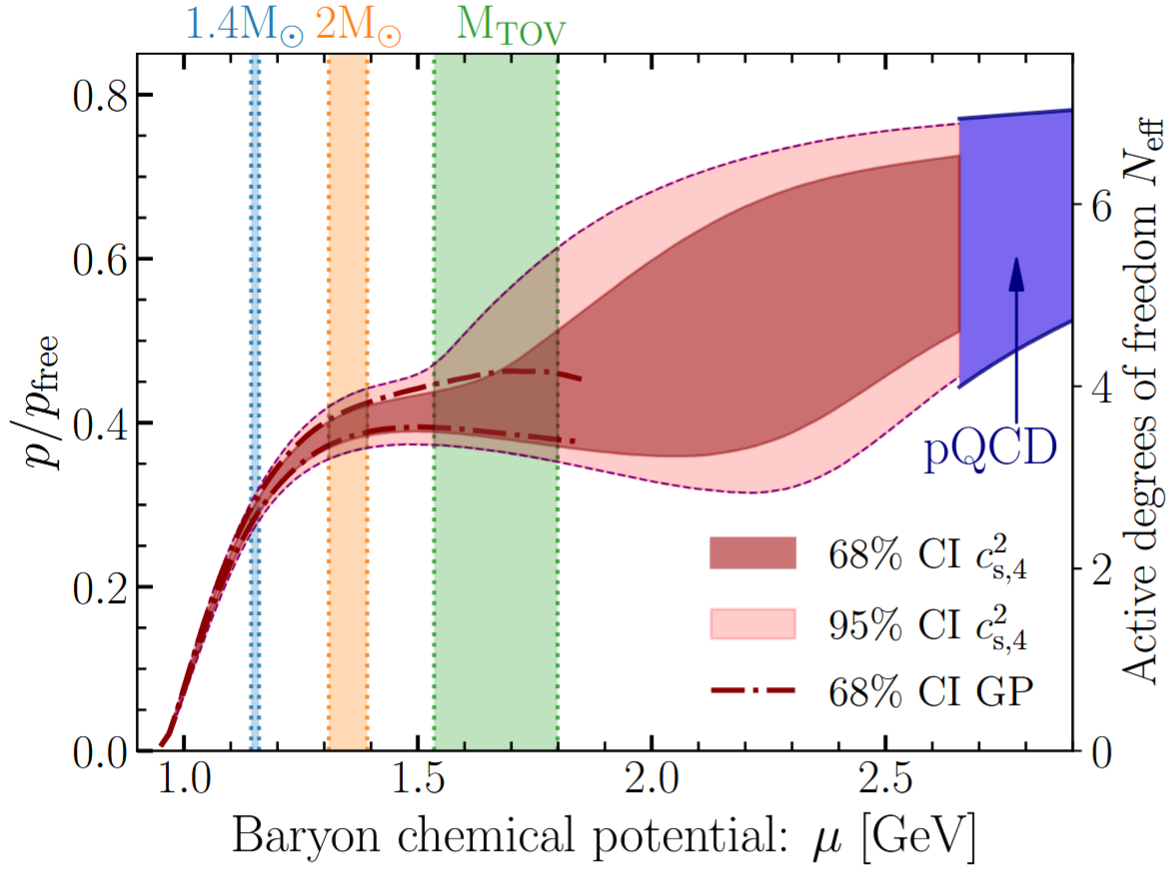}
    \caption{Left: the NLO, N2LO and N3LO pressures of cold and dense QM \cite{Gorda:2023mkk}, with the one unknown coefficient $c_0$ set to a value suggested by Bayesian arguments. Right: the model-independent pressure of cold beta-equilibrated NS matter as a function of baryon chemical potential \cite{Annala:2023cwx}, with the high-density pQCD constraint visible as the violet band on the right (see original paper for details).}
    \label{fig:press}
\end{figure}

In the conference proceedings at hand, we review recent progress towards the completion of the N3LO perturbative pressure of cold and dense QM --- a result expected to have a dramatic impact on NS-matter EoS inference (see Fig.~\ref{fig:press}). These advances originate from the fresh generalization of an established method of vacuum quantum field theory (QFT), the Loop Tree Duality \cite{Capatti:2019ypt}, to a thermal setting \cite{Navarrete:2024zgz}. There, it has already been used to evaluate a number of four-loop vacuum diagrams at vanishing temperature but nonzero quark chemical potentials $\mu_f$ \cite{Navarrete:2024zgz,Karkkainen:2025nkz} and very recently also a set of massive sum-integrals at nonzero temperatures \cite{Navarrete:2025yxy}. As argued in more detail below, the method's unique promise stems from its ability to reduce the evaluation of complicated multi-loop Feynman graphs to convergent multidimensional integrals, resulting from its power to systematically remove both ultraviolet (UV) and infrared (IR) divergences at the integrand level.

Another physical quantity we discuss in these proceedings is the bulk viscosity of (unpaired) QM, $\zeta$. It has been argued to play a key role in the dynamics of binary NS mergers \cite{Alford:2017rxf}, where it describes the dissipation of energy during radial density oscillations, leading to potentially substantial effects in the postmerger GW waveform. It, too, cannot be directly evaluated with \textit{ab-initio} tool at the relevant densities, although some simplifications occur due to the prominent role played by the electroweak interaction (see, e.g., \cite{CruzRojas:2024etx}), making bulk the evaluation of $\zeta$ considerably more straightforward than that of, e.g., its shear counterpart.

\section{Thermodynamic quantities}
\label{sec:thermo}

During the past more than 20 years, considerable effort has been placed in understanding the contribution of soft long-distance degrees of freedom to the equilibrium thermodynamics of deconfined QCD matter. Due to this work, heavily relying on the use of effective-field-theory (EFT) frameworks, all pressure contributions from momenta of order the electric and magnetic screening masses,\footnote{
Here, $N_c$ denotes the number of colors and $N_f$ the number of massless quark flavors.}
\begin{equation}
m_\text{E}=\frac{g}{\sqrt{3}} \sqrt{\left(N_c+\frac{N_f}{2}\right)T^2+\frac{3N_f}{2\pi^2} \mu_f^2}\,,\quad m_\text{M}\sim g^2T \, ,
\end{equation}
have by now been successfully determined up to and including $O(\alpha_s^3)$ in the strong coupling $\alpha_s=g^2/(4\pi)$. This is true for both hot \cite{Kajantie:2002wa,Vuorinen:2003fs} and cold but dense systems \cite{Gorda:2021znl,Gorda:2023mkk}, but one one crucial piece remains unknown in both cases: the contribution of hard momenta of order $T$ or $\mu_f$, available through the sum of all four-loop vacuum diagrams of the theory.

Unfortunately, the evaluation of Feynman integrals at nonzero temperatures or chemical potentials is very demanding business especially at high loop orders. 
Crucially, most automated tools of vacuum QFT become considerably less powerful in a thermal setting due to the effective breaking of Lorentz invariance to mere three-dimensional spatial rotations. 
Among the tools affected, the most notable is Integration-by-Parts (IBP) reduction \cite{Chetyrkin:1981qh,Laporta:2000dsw}, which is complicated by the appearance of extra "surface terms" when temporal momentum derivatives hit the Bose-Einstein or Fermi-Dirac distribution functions present at nonzero $T$ or $\mu$. 
This means that the space of Feynman integrals is expanded by new structures beyond those generated by the Feynman rules, which severely complicates the search of useful linear relations among the original integrals, and can only be avoided by restricting to purely spatial IBPs \cite{Nishimura:2012ee}.

For the reasons discussed above, progress with the necessary four-loop diagrams has so far relied on case-by-case computational tricks or methods that are either restricted to particularly simple topologies (see, e.g., the case of high-temperature $\phi^4$ theory discussed in \cite{Gynther:2007bw}) or built for the specific limit of $T=0$ and $\mu_f\neq 0$, such as cutting rules \cite{Ghisoiu:2016swa} or augmented IBP relations \cite{Osterman:2023tnt}. While relevant for the cold and dense limit of our interest, both of the latter methods unfortunately come with their own limitations:  the cutting rules suffer from the generation of spurious IR divergences and the appearance of nonstandard kinematic configurations in scattering amplitudes, while the augmented IBP relations are so far restricted to low loop orders \cite{Osterman:2023tnt} (see, however the  recent progress achieved in \cite{Nurmela:2025rtd}).

Fortunately, a subset of IBPs --- invariance under reparameterizations of loop momenta --- remains valid even in the thermal case \cite{Karkkainen:2025nkz,Navarrete:2024ruu}. For cold and dense QM, their systematic use was recently shown to reduce the original set of $O(10^5)$ four-loop integrals to $O(100)$ masters \cite{Karkkainen:2025nkz}, while in hot Yang-Mills theory, an even more impressive reduction down to $O(10)$ integrals was achieved in \cite{Navarrete:2024ruu}. In the former case, a local subtraction of all IR divergences has further transformed the completion of the full N3LO pressure into the evaluation of IR-finite (but UV-divergent) four-loop integrals, leaving an automated UV subtraction and a numerical evaluation of the finite integrals to be carried out. With this in mind, we next turn to a unique computational framework capable of meeting this challenge: the thermal Loop Tree Duality.

\subsection{Thermal Loop Tree Duality}
\label{sec:LTD}

The Loop Tree Duality (LTD) is an established method of vacuum QFT, where it is used for deriving locally finite three-dimensional momentum-space representations for scattering amplitudes \cite{Catani:2008xa,Capatti:2019ypt}. In this respect, it is closely related to several other approaches, including the Cross-Free-Family \cite{Capatti:2022mly}, Time-Ordered Perturbation Theory (see, e.g., \cite{SchwartzQFT}), and Flow-Oriented Perturbation Theory \cite{Borinsky:2022msp} approaches. Given the special treatment of temporal integrals in LTD, it is natural to expect the method to carry over to a thermal context with limited modifications, which prompted the authors of \cite{Navarrete:2024zgz} to apply it to the evaluation of vacuum Feynman diagrams at vanishing temperature but nonzero chemical potentials. 

According to conventions adopted in \cite{Navarrete:2024zgz,Karkkainen:2025nkz}, somewhat differing from the original vacuum notation, the thermal LTD procedure comprises four separate operations, all acting on a generic Feynman graph $\Gamma$ (possibly but not necessarily with external legs): (i) the subtraction of IR divergences; (ii) the (algorithmic) subtraction of UV divergences; (iii) the (algorithmic) derivation of a locally finite three-dimensional representation for the spatial momentum integrals; and (iv) the numerical evaluation of the remaining phase-space integrals.

Whether at nonzero temperature or density, step (i) requires knowledge of an EFT that applies to the soft modes of the theory under consideration. At high temperatures, the simplest EFTs are dimensionally reduced theories constructed for the Matsubara zero-modes \cite{Laine:2016hma}, while in the $T=0$ limit, this role is served by the Hard Thermal Loop (HTL) effective theory. The corresponding loop integrals arising in the EFTs capture the contribution of the IR modes in a resummed fashion and simultaneously serve as generators of local counterterms for full-theory graphs suffering from IR divergences.

In contrast, the subtraction of UV divergences in step (ii) applies a well-established automatable algorithm of vacuum QFT dating back to the early days of renormalization, the so-called "$R$-operation" \cite{Bogoliubov:1957gp,Hepp:1966eg,Zimmermann:1969jj}. 
Through the steps (i) and (ii), one arrives at an expression of the schematic form
\begin{equation}
\label{eq:R}
    \Gamma  = \bigg( \Gamma - \sum_j \text{CT}_j[\Gamma] \bigg) + \sum_j \text{CT}_j[\Gamma],
\end{equation}
where CT$_j[\Gamma]$ denotes the counterterms that remove all divergences at the integrand level, so that the expression inside the parentheses becomes locally finite in $D=4$ dimensions and can thus be integrated numerically.

Moving forward, step (iii) of the thermal LTD pipeline consists of two separate analytic computations, one for each of the two terms of Eq.~\eqref{eq:R}. For the finite expression in the parenthesis, one evaluates all temporal momentum integrals using the Residue theorem, which amounts to the step referred to as "LTD" in the vacuum context \cite{Catani:2008xa}. The remaining divergent integral in Eq.~\eqref{eq:R} is on the other hand either renormalized away using a suitable scheme (or resummed into EFT contributions in the case of IR divergences), or performed in its entirety within dimensional regularization in $D=4-2\varepsilon$ dimensions. The removal of poles in physical quantities then proceeds via standard vacuum renormalization together with resummations implemented within the EFT.

What remains at this point is step (iv), i.e.~the evaluation of the spatial integrations in the finite term, which typically involves some trivial angular integrals and a number of nontrivial integrations that are approached with Monte-Carlo methods. We refrain from a technical description of these details, but refer the interested reader to the recent PhD thesis of Kaapo Seppänen \cite{KSPhD}.

\begin{table}[t]
\centering
\begin{tabular}{||c|c|c|c|c|c|c|}
    \hline
    Diagram & $\varepsilon^{-2}$ & $\varepsilon^{-1}$ & $\varepsilon^0_\text{traditional}$ & $\varepsilon^0_\text{dLTD}$ & $N\,[10^6]$ & $[\mathrm{\mu s}]$ \\ 
    \hline \hline
    \raisebox{1.3\height}{$-\frac{1}{12}$}\includegraphics[height=0.9cm]{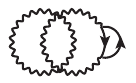} & \raisebox{1.3\height}{$0$} & \raisebox{1.3\height}{$\frac{3}{4}$} & \raisebox{1.3\height}{$12.375$} & \raisebox{1.2\height}{$12.36(4)$} & \raisebox{1.3\height}{$110$} & \raisebox{1.3\height}{$7.1$} \\
    \noalign{\hrule height.9pt} \noalign{\hrule height.9pt}
    \raisebox{1.3\height}{$-\frac{1}{8}$}\includegraphics[height=0.9cm]{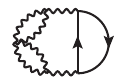} & \raisebox{1.3\height}{$\frac{27}{4}$} & \raisebox{1.3\height}{$\frac{189}{2}$} & \raisebox{1.3\height}{$716.38$} & \raisebox{1.2\height}{$716.32(7)$} & \raisebox{1.3\height}{$120$} & \raisebox{1.3\height}{$6.5$} \\
    \hline
\end{tabular}
\caption{First terms in the $\varepsilon=(3-d)/2$ expansion of the two diagrams considered in \cite{Karkkainen:2025nkz}. We omit the overall prefactor $d_AN_fC_A^2g^6d(d-1)\mu^4/(4\pi)^8$ and set the $\overline{\text{MS}}$ renormalization scale to the value $\bar{\Lambda}=2\mu_\text{B}/3$. The symbol $N$ indicates the number of Monte Carlo samples needed and $[\mathrm{\mu s}]$ the time (in microseconds) spent on one sample by a single CPU core, while dLTD denotes "dense LTD".}
\label{table1}
\end{table}

In the original paper that generalized the LTD framework to a thermal setting \cite{Navarrete:2024zgz}, the authors applied the method to one particularly complicated four-loop vacuum diagram at vanishing temperature but nonzero quark chemical potentials. Following the evaluation of this intrinsically IR- and UV-finite entity, responsible for the leading perturbative difference between the pressures of full and phase-quenched QCD \cite{Moore:2023glb}, the method was next used in a proof-of-principle study of the N3LO pressure of cold and dense QM in  \cite{Karkkainen:2025nkz}. This article not only demonstrated the gauge invariance and IR finiteness of the pressure at this order, but also evaluated two non-factorizing four-loop graphs in two independent manners, using both LTD and standard methods. As demonstrated in Table \ref{table1}, the LTD results agree with the semianalytic ones to the desired accuracy, with the numerical computation involved in the LTD calculation only taking a few minutes on a single CPU. This showcases the applicability of the LTD framework at the four-loop level and demonstrates its efficient scalability to the most demanding integrals, for which standard methods are not available.

In a very recent exploratory study, the thermal LTD framework was further extended to nonzero temperatures and masses in the perturbative evaluation of the thermal effective potential of a simple toy model exhibiting a strong Electroweak phase transition \cite{Navarrete:2025yxy}. 
Analogously to the case of chemical potentials, the introduction of a nonzero temperature amounts to a very benign qualitative change in the LTD pipeline, with the main exception being found in step (i), the IR subtractions. 
Here, the realm of high temperatures, however, turns out more straightforward than the cold and dense limit. 
As explained earlier, at sufficiently high $T$, IR physics can namely be described using dimensionally reduced effective theories, such as Electrostatic QCD (EQCD) \cite{Appelquist:1981vg,Braaten:1995jr,Kajantie:1995dw}.
The lower dimensionality of these theories leads to considerably simpler integrand-level matchings than those encountered with the HTL effective theory, needed in the small-$T$ limit \cite{Karkkainen:2025nkz}.
Altogether, these generalizations show great potential in extending state-of-the-art QCD results to higher loop orders and the most general case of nonzero temperatures, densities, and quark masses \cite{thermalLTDinprep}, also under consideration at the moment. 

Finally, let us briefly comment the role of quark-pairing corrections to the perturbative expansion of the QM EoS. 
The reason this topic has not been extensively addressed in the literature is that the (additive) pairing contribution to the pressure is expected to be suppressed by a factor of $\Delta^2/\mu_\text{B}^2$ with respect to the unpaired part, with $\Delta \lesssim 100$ MeV standing for the the pairing gap (see, e.g., discussion in \cite{Kurkela:2009gj}). 
Taken at face value, this leads to very small numerical effects at densities where the perturbative pressure is currently applied in NS-matter EoS inference, but there are a few caveats involved in this statement. 
First, should the completion of the N3LO pressure improve the accuracy of the result enough, so that the result becomes applicable down to $\mu_\text{B}\approx 2$ GeV, the pairing contributions may become substantial and need to be revisited. 
Similarly, a new approach to the pressure of cold and dense QM, involving phase-quenched lattice simulations (see, e.g., \cite{Moore:2023glb}), is believed to involve substantial pairing contributions at small temperatures \cite{Fujimoto:2023mvc}, which necessitates a detailed study of these effects unless one restricts attention to temperatures large enough so that the pairing contributions can be altogether neglected \cite{Gorda:2025cwu}.



\section{Bulk viscosity}
\label{sec:bulk}

As briefly discussed in Sec.~\ref{sec:intro} above, realistic numerical simulations of binary NS mergers need to account for more than just the thermodynamic properties of the medium, namely bulk viscous dissipation in the system \cite{Alford:2017rxf}. In fact, it has been speculated that such effects might offer a potential means to detect the generation of large amounts of deconfined matter during NS mergers should the bulk viscosities of HM and QM behave in sufficiently differing ways (see, e.g., \cite{Most:2021zvc,Chabanov:2023blf,Haber} for related discussions).

\begin{figure}
    \centering \includegraphics[width=0.6\textwidth]{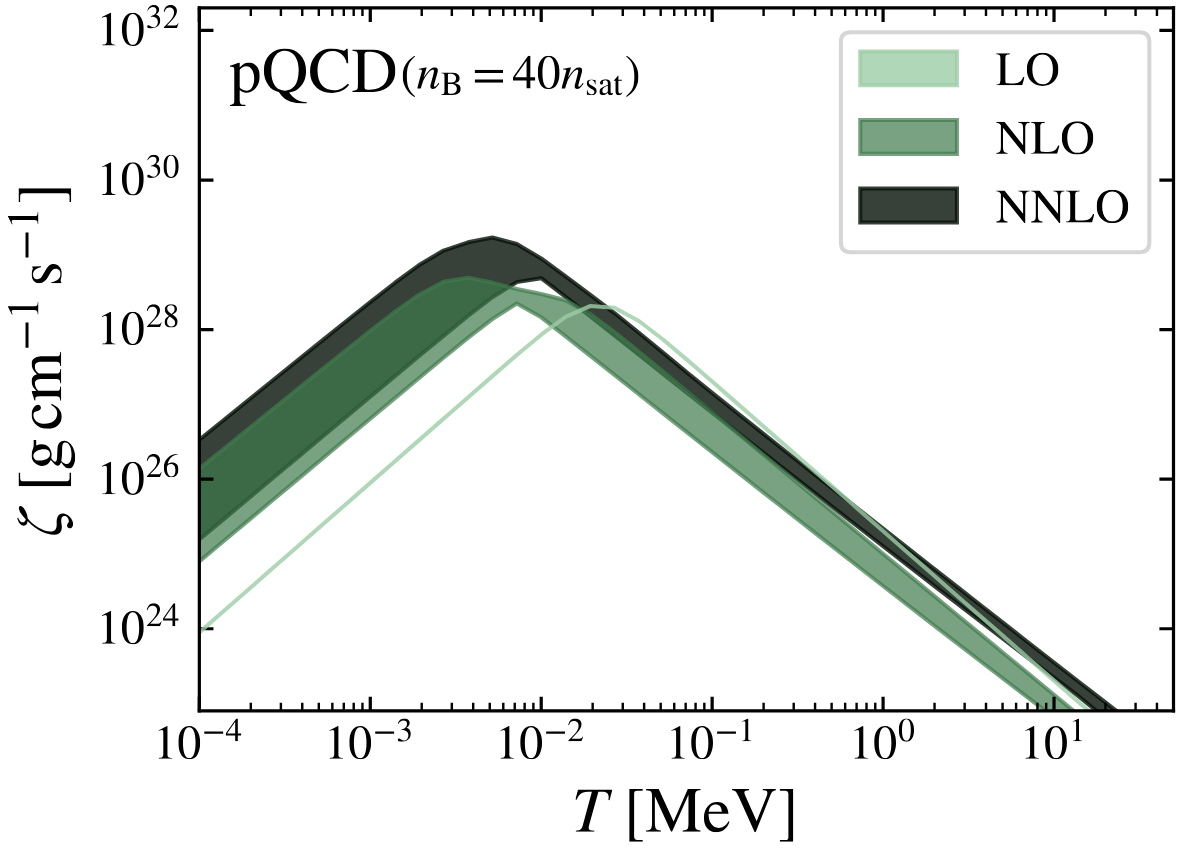}
    \caption{The non-leptonic contributions to the LO, NLO, and N2LO bulk viscosity of unpaired QM, taken from the recent \cite{CruzRojas:2024etx}.}
    \label{fig:bulk}
\end{figure}

The bulk viscosity of dense QCD matter differs from other transport properties not only by its phenomenological importance but also due to the physical mechanism behind it. As reviewed in detail in the Supplemental Material of \cite{CruzRojas:2024etx}, $\zeta$ arises from a subtle interplay between the strong and electroweak interactions, where the system tries to maintain chemical equilibrium while periodic radial pulsations arising from the system's macroscopic dynamics keep perturbing it. In QM, the final expression for the quantity,
\begin{eqnarray}
\zeta&=&\frac{\lambda_1 A_1^2}{\omega^2+(\lambda_1) C_1^2},
\end{eqnarray}
contains two types of microphysical inputs: the rates of the relevant electroweak processes, either purely non-leptonic or both non-leptonic and semi\-leptonic in nature (here $\lambda_1$), and thermodynamic coefficients taking the form of linear combinations of various quark number densities and susceptibilities (here $A_1$ and $C_1$).

To improve the description of the bulk viscosity of unpaired QM (see, e.g., \cite{Sad:2007afd,Alford:2013pma}), it is crucial to improve both the rates and the thermodynamic quantities entering the result. Here, it is interesting to note that the thermodynamic inputs can be obtained with methods perfectly analogous to those used for the QM EoS, with the only subtlety related to the need to keep the strange quark mass nonzero. This, too, is however only a minor complication given the mass expansion scheme of \cite{Gorda:2021gha} that allows treating $m_s$ as a perturbation and expanding thermodynamic quantities in its integer powers.

Recently, the bulk viscosity of unpaired QM was extended to an unprecedented N2LO accuracy in a series of works, of which \cite{CruzRojas:2024etx} only accounted for the non-leptonic process $u+d \leftrightarrow u+s$, a valid approximation at very small temperatures, while \cite{Hernandez:2025zxw,Sappi:2025tcx} also added the effects of semileptonic processes using a two-component "Burgers' fluid" description. As seen from Fig.~\ref{fig:bulk}, taken from \cite{CruzRojas:2024etx}, the results display a strong sensitivity to the perturbative order even at baryon densities of order $40n_s$, highlighting the need to drive the calculations towards higher loop levels and possibly consider approches such as holography \cite{Hoyos:2020hmq}. Here, the thermal LTD framework will undoubtedly be an invaluable asset, but the efforts must be complemented with $O(\alpha_s)$ corrections to the electroweak rates, currently under investigation \cite{rates}.

Finally, we note that the bulk viscosity is expected to be highly sensitive to quark pairing (see, e.g., the discussion in \cite{Schmitt:2017efp}), and to this end, all unpaired results for the quantity should merely be viewed as starting points in the investigation of true bulk viscous effects. As this topic, however, extends rather far from the scope of these conference proceedings, we refrain from a more extensive discussion of the issue here.




\section{Future directions}

Figuring out the composition and properties of neutron-star interiors with first-principles methods is a notoriously challenging problem, with the strongly coupled nature of the system invalidating naive weak-coupling approaches and the Sign Problem preventing a straightforward use of lattice methods. During the past decade, systematic work on the model-independent inference of the NS-matter EoS, utilizing theoretical \textit{ab-initio} limits and robust astrophysical observations, has nevertheless resulted in significant advances. In addition to increasingly accurate EoS results, this approach has led to first-ever evidence for the likely presence of quark matter inside massive NSs \cite{Annala:2019puf,Annala:2023cwx}, where the perturbative pressure of cold and dense QM has played a very important role (see, e.g., \cite{Komoltsev:2021jzg}). 

In the conference proceedings at hand, we have reviewed recent advances in the perturbative description of the thermodynamic and transport properties of dense QM. A particularly important breakthrough can be found in the recent generalization of the Loop Tree Duality method of vacuum quantum field theory to a thermal setting, recently achieved in \cite{Navarrete:2024zgz,Karkkainen:2025nkz,Navarrete:2025yxy}. 
The strength of the thermal LTD approach is rooted in its ability to reduce complicated perturbative calculations to the numerical evaluation of convergent multidimensional momentum integrals. This property makes it blind to many details that severely complicate (semi)analytic approaches, such as the presence of nonzero quark masses, small but nonzero temperatures, and multiple independent chemical potentials. In addition to the completion of the N3LO pressure of cold QM, near-future applications of the method include at least the extension of this result to nonzero temperatures, with the pressure of hot Yang-Mills-theory as an interesting limiting case, and a systematic exploration of a wide class of Beyond-the-Standard-Model scenarios expected to feature strong Electroweak phase transitions. 








\end{document}